\documentstyle[preprint,aps]{revtex}

\date{April 16, 1997}

\title{Density of Superfluid Helium Droplets}

\author{Jan Harms, J.Peter Toennies} 

\address{Max-Planck-Institut f\"ur Str\"omungsforschung, 37073 G\"ottingen,
  Germany}

\author{Franco Dalfovo} 

\address{Dipartimento di Fisica, Universit\`a di Trento, 38050 Povo, Italy}

\begin{document}

\maketitle

\begin{abstract}
  The classical integral cross sections of large superfluid $\rm ^4He_{_N}$
  droplets and the number of atoms in the droplets ($N=10^3-10^4$) have been
  measured in molecular beam scattering experiments. These measurements are
  found to be in good agreement with the cross sections predicted from density
  functional calculations of the radial density distributions with a 10--90\%
  surface thickness of 5.7\AA{}. By using a simple model for the density
  profile of the droplets a thickness of about 6--8\AA{} is extracted directly
  from the data.
\end{abstract}

%##########################################################################

\section{Introduction}
\label{sec:introduction}
The study of the properties of $^4$He clusters is currently an active area of
theoretical and experimental research \cite{W94,T90}. The interest is based on
the fact that $^4$He clusters provide an easily accessible example of a finite
size quantum system of strongly correlated particles. One of the primary aims
of He cluster research is to search for manifestations of superfluidity in
mesoscopic systems.  Recent spectroscopic studies of glyoxal molecules
embedded in $^4$He droplets with about 5000 atoms produced in free jet
expansions provide the first evidence that these droplets are indeed
superfluid \cite{HMTV96}. The internal temperature of these droplets has also
been measured spectroscopically to be about $0.4~K$ \cite{HMTV95} in agreement
with theory \cite{BS90}.

Even though there has been notable progress in theoretical work on He clusters
during the last years the agreement between theory and experiments is not yet
entirely satisfactory \cite{W94}. The implementation of {\sl ab initio}
calculations still remains difficult for inhomogeneous systems with more than
a few hundred atoms, while phenomenological theories, which work well for
describing macroscopic properties, are often not adequate for systems of
finite size. A major progress to fill this gap has been made in the
development and use of density functional (DF) theory \cite{DLP95}.

The understanding of the density profiles in the surface region of a quantum
fluid has long been considered a very important basic problem \cite{ED78}. For
this reason significant effort has gone into the experimental determination of
the thickness of the bulk liquid-vapor $^4$He interface \cite{ED78}. At least
three different experimental methods have been used to measure the surface
thickness of bulk liquid He. The first involved atomic scattering experiments
\cite{NEM83} yielding an interfacial width (10--90\%) of about 4\AA{}. A
surface thickness of about 9\AA{} was measured with an ellipsometric method
\cite{O89}. Finally, x-ray measurements give a surface thickness of the order
of 8\AA{}, providing also information on the shape of the interface
\cite{LRP92}. In droplets the surface atoms represent a significant fraction
of the total number of atoms.  For example, in a droplet with 5000 atoms about
one half of the atoms are located in the surface region.  Thus the physical
properties of $^4$He droplets are strongly influenced by the surface.
Moreover, the surface region is particularly interesting because recent
calculations predict that the Bose-condensed fraction, which in bulk $^4$He is
about 10\%, approaches 100\% in regions where the density is still within 10\%
of the bulk density \cite{GS95}.

In this paper the liquid-vapor interface of large $^4$He droplets
($N=10^3-10^4$) is investigated by a new method. The experiment consists in
measuring both the integral cross section of the droplets and also their
number size distributions. From the integral cross section and the average
number of atoms an effective droplet volume as well as an average density is
determined. The results are compared with theoretical predictions. Although a
large number of calculations have been performed for the density profiles of
small $^4$He clusters ($N < 10^3$)
\cite{W94,CDL96,SKC89,CK95,BW95,PZP83,PPW86,ST87} no calculations for the
large droplets investigated here have been reported so far.  In the present
work DF theory is used to calculate the density profiles of large $^4$He
droplets. The density functional used was recently introduced by the
Orsay-Trento collaboration \cite{DLP95}. This functional provides density
profiles and energies \cite{CDL96} very close to the results of {\sl ab
  initio} Monte Carlo calculations \cite{SKC89,CK95,BW95} in the case of small
clusters and has the advantage that it can easily be applied to droplets with
several thousands of atoms. These density profiles provide for a direct
comparison with the experimental data and good agreement is found within the
experimental errors.

The droplet beam scattering apparatus is described in the following section.
The procedure used to measure the number of atoms in the droplets and their
classical integral cross sections are described in Section
\ref{sec:droplet_sizes} and \ref{sec:integral_cross_sections}. In Section
\ref{sec:mean_droplet_densities} the experimental results are presented and in
Section \ref{sec:comparison_with_theory} they are compared with the DF
predictions.  A final discussion is given in Section \ref{sec:discussion}.

%##########################################################################

\section{Experimental}
\label{sec:experimental}
Since the experimental methods and the apparatus used in this work have
already been described in some detail in previous publications
\cite{LST93,LST95} the experimental procedure will be only briefly described
here.  The $^4$He droplet beam is produced by a free jet expansion of $^4$He
from a stagnation source chamber at a high pressure and at low temperature
(typically $P_0=5-100~bar$, $T_0=4-30~K$) through a thin walled ($20~{\mu}m$)
nozzle with $5\pm1~{\mu}m$ diameter. After the droplets have passed the
skimmer and several differential pumping stages they are scattered by a
secondary beam produced in another free jet expansion beam source which
crosses the droplet beam at an angle of $40^\circ$. The scattered droplets are
detected with a homemade electron impact ionizer optimized for a high
ionization efficiency followed by a magnetic mass spectrometer. The detector
can be rotated around the scattering region in the plane of the two beams. To
avoid capture collisions with the residual gas the background non-helium
pressures in the differential pumping stages between source and scattering
chamber were kept below $10^{-7}~mbar$.

%##########################################################################

\subsection{The Number of Atoms in the Droplets}
\label{sec:droplet_sizes}
The atom number distribution $P(N)$ of the helium droplets is determined from
the angular distribution resulting from scattering by the secondary beam.  A
small portion (5--10\%) of the droplets is deflected by the momentum transfer
imparted by single collisions of the secondary beam gas atoms, most of which
are captured by the droplets \cite{LST95}.

The measurements described in the present study have been carried out at
source pressures of $P_0=40~bar$ and $P_0=80~bar$ and nozzle temperatures
between $T_0=13~K$ and $T_0=26~K$. For these conditions the droplet velocity
distributions are very narrow ($\frac{\Delta v}{v} \approx 2\%$) with well
known mean speeds \cite{BTN91}. The secondary beam gases Ar and Kr were
expanded from stagnation temperatures of about $300~K$ and several hundred
millibar of stagnation pressure from a $40~{\mu}m$ diameter nozzle.  The
secondary beam contains no appreciable amount of dimers or larger clusters,
and has a narrow velocity distribution ($\frac{\Delta v}{v} \approx 20\%$).

The angular distribution of the deflected droplet beam is measured by rotating
the detector in small angular increments (typically $300~\mu{}rad$) around the
scattering center. Due to the large mass of the droplets the deflection angles
are very small and a high angular resolution is necessary. This is achieved by
collimating the incident beam with a $50~{\mu}m$ slit in front of the
scattering center and a $25~{\mu}m$ slit in front of the detector. The width
of the expansion zone and the broadening due to collisions with the residual
background gas (mostly helium) leads to a FWHM of the beam profile of about
$1.5~mrad$ which defines the effective angular resolution.

Fig.~\ref{fig:distr} shows three typical measured angular distributions with
and without a flag in front of the secondary beam \cite{LST93}. The small
difference between the two signals is attributed to droplets which were
deflected after capturing a secondary beam atom. Assuming complete momentum
transfer the angle of deflection is directly dependent on the number of atoms
in droplet.  As is discussed in more detail below, the present experiments
give further justification for the hypothesis of complete momentum transfer.
The measured droplet atom number distributions can be very well fitted with a
log-normal-distribution \cite{LST93,LST95}:
\begin{equation} 
P(N)  = 
  \frac{1}{\sqrt{ 2 \pi } N \delta } \: \exp\left[- \frac{ (\ln N - 
  \mu)^2 }{ 2 \delta^2}
  \right] ,
\label{eq:lognorm} 
\end{equation}
where the mean number of atoms $\overline{N}$ and the width (FWHM) are
\begin{equation}
  \overline{N} =  \exp \left(  \mu + \frac{\delta^2}{2} \right)  
\end{equation}
and
\begin{equation}
  \Delta N_{\frac{1}{2}} =  \exp\left(\mu - \delta^2 + \delta \sqrt{2 \ln
      2} \right) - \exp \left( \mu - \delta^2 - \delta \sqrt{2 \ln 2} \right)
\; ,
\end{equation}
respectively. The FWHM is comparable with $\overline{N}$.
Table~\ref{tab:exp_tab} lists the results for $\overline{N}$, $\Delta
N_{\frac{1}{2}}$, $\delta$ and $\mu$ , measured using Ar and Kr as deflecting
atoms for a wide range of different $^4$He source pressures and temperatures.
The values of $\overline{N}$ and $\Delta N_{\frac{1}{2}}$ obtained using Ar or
Kr for identical $^4$He source conditions do not always agree exactly, since
they depend sensitively on the nozzle-diameter and on the nozzle-skimmer
distance which was optimized for each experiment and differed slightly.  In
particular the mean droplet sizes for $P_0=80~bar$ with krypton as secondary
gas are somewhat smaller by about 30\% since the nozzle used in these
experiments had a diameter of only about $4~{\mu}m$ as estimated from the
total gas flow into the nozzle chamber. These deviations however have no
appreciable effect on the average densities. The latter depend on the number
of atoms in the droplets and the integral cross sections, both of which are
determined in the same experiment under identical source conditions (see
below).

%##########################################################################

\subsection{Integral Cross Sections}
\label{sec:integral_cross_sections}
The integral cross sections of the droplets are determined by measuring the
attenuation of the forward peak in the deflection pattern (see
Fig.~\ref{fig:distr}). The attenuation is related to the integral cross
section $\overline{\sigma}$ of the droplets according to Beer's law
\cite{LST95},
\begin{equation}
  \frac{I}{I_0} = \exp \left[
        - \frac{\overline{\sigma} \: n_{\text{sec}} L_{\text{eff}} \: 
      v_{\text{rel}} \:   F_{\text{a0}}}
   {v_{\text{drop}}} \right] \; ,
\label{eq:beer}
\end{equation}
where $I$ and $I_0$ are the intensities of the $^4$He droplet beam with and
without attenuation, $n_{\text{sec}}$ is the number density of the secondary
beam gas atoms in the scattering center and $L_{\text{eff}}$ is the effective
length of the scattering volume.  Moreover, $v_{\text{drop}}$ is the speed of
the droplets and $v_{\text{rel}}$ the relative collision velocity.
$F_{\text{a0}}$ takes account of the smearing resulting from the velocity
distributions of the two nozzle beams \cite{BHK62} and leads to a correction
smaller than about 1\%.

The product of the density in the scattering center and the effective length
of the scattering volume $(n_{\text{sec}}\cdot L_{\text{eff}})$ was calibrated
to within an error of approximately 5\% by measuring the attenuation of a
nearly monoenergetic helium atomic beam, for which the integral cross section
can be accurately calculated quantum mechanically from the well established
interaction potential \cite{TT86}. The density was calibrated three times with
krypton and two times with argon at different collision energies and there was
no evidence of systematic errors. The values of the absolute integral cross
sections of the droplets are also listed in Table~\ref{tab:exp_tab}.

The interpretation of the integral cross sections in Eq.~(\ref{eq:beer}) is
straightforward only in the ideal case of an infinite angular resolution. In
this case one gets the quantum mechanical integral cross section, which
includes the forward peaked diffraction part. In the present case of finite
angular resolution the measured integral cross section is smaller since a
fraction of the forward diffraction part of the differential cross section is
not included. In order to estimate this correction, the quantum mechanical
differential cross section for elastic scattering was calculated using a
partial wave expansion method \cite{EST88}.  The calculations, which are
described in detail in Appendix~\ref{app:cs}, indicate that the contribution
of the diffraction part to the elastic cross section is negligible with the
angular resolution of our apparatus.  Thus the measured integral cross section
$\sigma_{\text{app}}^{\text{int}}$ turns out to be very close to the classical
cross section $\sigma_{\text{class}}$.  The two cross sections agree to within
1\% for droplets with 1000 atoms and for larger droplets the agreement is even
better.

The finite width of the beam-profile also has the effect that the deflections
of the largest droplets in the atom number distributions are too small to lead
to a measurable attenuation of the beam. This error however can be estimated
since the atom number distributions of the droplets and the beam-profiles
$S(\Theta)$ are both well known from the measurements. The calculation yields
a correction of about 1\% for mean atom numbers of $10^3$ and of about 10\%
for droplets with $10^4$ atoms.

%##########################################################################

\subsection{Mean Droplet Densities}
\label{sec:mean_droplet_densities}
Measured values of the integral cross section, $\overline{\sigma}$, as a
function of the mean atom numbers $\overline{N}$ are reported in
Fig.~\ref{fig:cs}. The values obtained with argon and krypton as secondary gas
fall on a common curve and so we can conclude that the integral cross section
is independent of the nature of the secondary beam, within the estimated
accuracy. The overall experimental errors are estimated to be about 5\%
and result mainly from the uncertainty in the determination of the density of
the secondary beam atoms in the scattering volume. 

From the measured cross
section an effective mean density $\overline{\rho}$, defined as the density of
a uniform sphere with a sharp step edge (liquid drop model) having the same
classical integral cross section, is determined
\begin{equation}
 \overline{\rho} = 
  \frac{3}{4} \sqrt{\frac{\pi}{\overline{\sigma}^{^{\;3}}}}\;\overline{N} \;. 
\end{equation}
The values of $\overline{\rho}$, normalized to the bulk helium density
$\rho_{\text{bulk}}=0.0218$ \AA$^{-3}$, are given in the last column of
Table~\ref{tab:exp_tab} and the integral cross sections for spheres of
different density are shown in Fig.~\ref{fig:cs} as dashed lines.  Here the
average integral cross section $\overline{\sigma}$ is calculated from the
measured log-normal atom number distributions $P(N)$ by means of the following
equation:
\begin{equation}
  \overline{\sigma(\overline{N})} = \int_0^\infty P(N) \: \pi^{\frac{1}{3}}\:
  \left(\frac{3 N}{4 \overline{\rho}}\right) ^\frac{2}{3} \: dN \; .
\label{eq:sig}
\end{equation} 
As can be seen, the experimental effective droplet density goes from about
$0.5 \cdot \rho_{\text{bulk}}$, for droplets with $10^3$ atoms, to $0.7 \cdot
\rho_{\text{bulk}}$ for droplets with $10^4$ atoms. This trend is consistent
with the fact that a significant part of the atoms are in the outer surface
region, where the density is less than the bulk density.

%##########################################################################

\section{Comparison with theory} 
\label{sec:comparison_with_theory}
For comparison with the above experimental results the integral cross sections
were calculated using density functional theory.  These calculations were
carried out with the improved density functional recently developed
\cite{DLP95} for the accurate calculation of $^4$He droplet properties. The
energy of the system is assumed to be a functional of the complex function
$\Psi$:
\begin{equation}
   E \ = \ \int \! d{\bf r} \ {\cal H} [\Psi,\Psi^{*}] \ ,
\label{eq:edf}
\end{equation}
\begin{equation}
  \Psi({\bf r},t) \ = \ \Phi({\bf r},t) 
       \exp \left( \frac{i}{\hbar} S({\bf r},t)  \right) \ .
\end{equation}
The real function $\Phi$ is related to the diagonal one-body density by
$\rho=\Phi^2$, while the phase $S$ fixes the velocity through the relation 
${\bf v} = (1/m) \nabla S$. In the calculation of the  ground state, only
states with zero velocity are considered, so that the energy is simply a
functional of the one body density $\rho({\bf r})$.
A natural representation is given by
\begin{equation}
E \ = \ \int \! d{\bf r} \ {\cal H} [\rho] \ =  \  E_c[\rho] \ + \
\int \! d{\bf r}  \ {\hbar^2 \over 2m}  ( \nabla \sqrt{\rho})^2  \ \ \ ,
\label{eq:ec}
\end{equation}
where the second term on the r.h.s. is a quantum pressure, corresponding to
the kinetic energy of a Bose gas of non-uniform density. The quantity
$E_c[\rho]$ is a {\sl correlation energy}, which  incorporates the effects of
dynamic correlations resulting from the interactions between the individual He
atoms.  The equilibrium configurations are obtained by minimizing the energy
with respect to the density. This leads to the Hartree-type equation
\begin{equation}
\left\{ - {\hbar^2 \over 2m} \nabla^2  + U [\rho, {\bf r}] \right\}
\sqrt{\rho({\bf r})} \ = \ \mu_4 \sqrt{\rho({\bf r})} \ \ \ ,
\label{eq:hartree}
\end{equation}
where $U[\rho, {\bf r}] \equiv \delta E_c /\delta \rho({\bf r})$ acts as a
mean field, while the chemical potential $\mu_4$ is introduced in order to
ensure the proper normalization of the density to a fixed number of particles
$N$. The density dependence of the correlation energy is parameterized in a
phenomenological way, by choosing a functional form compatible with basic
physical requirements and fixing a few parameters in accordance with the known
properties of the bulk liquid. The detailed form of $E_c$ is given in
Appendix~\ref{app:functional}. The DF theory is particularly suitable for the
calculation of the density profiles of relatively large droplets. In fact, the
density functional of Ref.~\cite{DLP95} is accurate, by construction, in the
limit of the uniform liquid. It also has been tested in the opposite limit of
small clusters with 20 -- 70 atoms \cite{CDL96} for which it provides density
profiles in close agreement with {\it ab initio} Monte Carlo calculations.

The density profiles shown in the upper part of Fig.~\ref{fig:profiles} are
obtained by solving Eq.~(\ref{eq:hartree}) for $N=10^3$ to $10^4$.  In order
to estimate the contribution to the effective integral cross section
$\sigma_{\text{eff}}$ from the outer region of the theoretical profiles, the
transmission coefficient $T(b)$ of the droplets is calculated as a function of
the impact parameter $b$ for argon and krypton using Beer's law (see the inset
in Fig.~\ref{fig:profiles}b). For this the trajectory of relative motion is
integrated along a straight path ${\bf z}$, which is a good approximation in
the outer region where the density is about 1\% of the core density (see
below):
\begin{equation}
        T(b) = \exp \left(
         -\int_{{\bf z}(b)}
        \sigma(E_{\text{rel}}) \: \rho({\bf z}) \: d{\bf z} \; \right).    
\label{eq:tvonr}
\end{equation}
The atom-atom integral cross sections $\sigma(E_{\text{rel}})$ were again
calculated using the method given in Appendix~\ref{app:cs}, with the relative
velocities of the experiment. Here, the more realistic Tang-Toennies
potentials \cite{TT86} are used in place of the Lennard-Jones(12,6)
potentials. The calculated transmission $T(b)$ of the droplets is illustrated
in Fig.~\ref{fig:profiles}b. The differences between argon and krypton are
less than 1\% and can therefore be neglected. The transmission for densities
larger than about 10\% of the central density is almost zero, so that the
cross section is largely determined by the outer region of the droplet.  Since
the transmission rises very steeply the effective droplet-border is rather
sharp and corresponds to a radius ($R_{\text{eff}}$) where the density has
fallen to approximately 1\%.  The integral cross section of the droplets is
then calculated with the following equation:
\begin{equation}
  \sigma_{\text{eff}} = 2 \pi \int_0^\infty \left(1-T(b) \right) b \: db \;. 
\label{eq:trans}
\end{equation}
In a subsequent step the cross sections are averaged over the distribution in
the number of atoms $P(N)$. The calculated results for the effective radii
$R_{\text{eff}} = \sqrt{\sigma_{\text{eff}} / \pi}$, $\sigma_{\text{eff}}$,
the average relative densities $\overline{\rho} / \rho_{\text{bulk}}$ and the
10--90\% thickness $t$ are reported in Table~\ref{tab:tab_dfc}. The average
relative densities $\overline{\rho} / \rho_{\text{bulk}}$ in Fig.~\ref{fig:cs}
(filled diamonds) lie between 0.64 and 0.8 and are somewhat larger than the
experimental values by about 10--25\%.  This discrepancy suggests that the
theoretical surface thicknesses are somewhat smaller than the experimental
values.

A simple model calculation can be used to check the effect of the surface
thickness on the integral cross section. Since all theoretical calculations
predict the core density of these large droplets to be equal to the bulk
density, the density profile can be rather realistically described by a simple
analytic function \cite{TGR85,J92}:
\begin{equation}
  \rho(r) = \frac{\rho_{\text{bulk}}}{2} \left(1 - \tanh \left(2
  \frac{r-R}{g}\right)\right)  \; ,
\label{eq:rho_r}
\end{equation}
where $R$ indicates the point where the density is reduced to 50\% of the
central density and $g$ is a parameter controlling the surface thickness.  By
relating the density profile to the integral cross section using the same
procedure described in Section \ref{sec:comparison_with_theory} an
experimental surface thickness is estimated by fitting the experimental cross
section for the different droplets.  The resulting values of $t$ are shown in
Fig.~\ref{fig:exp_prof}. We obtain a mean 10--90\% thickness of $6.4 \pm 1.3$
\AA, which is somewhat larger than the results of the DF calculation (Table
\ref{tab:tab_dfc}) but still within the error limits. One has to stress here
that the results are dependent on the model used for $\rho(r)$; in particular,
it assumes a symmetric density profile.  Most of the predicted density
profiles are however slightly asymmetric, with a steeper slope in the outer
part of the surface. By using an asymmetric profile, with an asymmetry similar
to the one of the DF profiles in Fig.~\ref{fig:profiles}, the estimated
surface thickness is about 1--2\AA\ larger than the values shown in
Fig.~\ref{fig:exp_prof}. This would increase the discrepancy between the
experimental estimates and the theoretical surface thickness.

%##########################################################################

\section{Discussion} 
\label{sec:discussion}

As discussed above it appears as if the experimental surface thickness is
somewhat larger than the values predicted by the DF theory. This is indicated
by the 10--25\% differences in the average densities in Fig.~\ref{fig:cs} and
by the larger widths obtained using model profiles similar to
Eq.(\ref{eq:rho_r}). Since these differences are greater than the experimental
uncertainties of 5\%, we consider here possible sources of error.

The possibility that the measured droplet atom numbers $\overline{N}$ are too
small, which would shift the mean effective density to smaller values, seems
unlikely. This would require a momentum transfer in the deflection experiment
even larger than the momentum ${\bf p}_{\text{sec}}$ of the impinging
secondary atom.  For elastic backscattering of the atoms in central collisions
from a hard sphere the momentum transfer would indeed be $2{\bf
  p}_{\text{sec}}$. The momentum transfer averaged over all impact parameters
for elastic scattering from a hard sphere is, however, equal to the momentum
of the incoming particle ${\bf p}_{\text{sec}}$. Thus even in the unlikely
case of elastic recoil scattering, which in view of the large mass of the
argon and krypton atom compared to that of the He atom seems very unlikely and
is not even expected for the scattering from a liquid helium surface
\cite{NEM83}, the same results would still hold.  Recoil scattering would also
be in contradiction to the observation that the secondary gas is captured by
the He droplets \cite {LST93,LST95}. Another possibility is a backward
directed vaporization of helium induced by the impinging secondary gas atom,
as proposed by Gspann \cite{G81}. According to ordinary fluid dynamics this
would also seem implausible, since the impact speeds are greater than the
velocity of first sound of about $240~m\!/\!s$ \cite{G81} and therefore a
conical Mach shock should be created which degenerates in a spherical sound
wave. This sound wave would more probable induce vaporization in forward
direction which would tend to shift our results to greater droplet atom
numbers $\overline{N}$ and would therefore increase the discrepancy with the
theoretical results.  

The comparison between measured cross sections and DF
calculations is affected by at least two possible sources of problems.  The
first concerns the form and extent of the density profiles, and the second the
overall shape of the droplets.  The effect of different density profiles was
already discussed in the previous section and has a noticeable effect. The
10--90\% surface thickness of the density profiles from the DF calculations is
a $t \cong 5.7$ \AA.  This value is slightly less than other estimates. For
example, previous DF calculations for a planar free surface \cite{GCB92} gave
$t \cong 7$ \AA{}, while from {\it ab initio} calculations on small clusters
\cite{SKC89,CK95,BW95,PZP83,PPW86,ST87} values between 6 and 7 \AA{} are
predicted.  Some experimental and theoretical results for the surface
thickness of the bulk liquid and droplets are summarized in
Table~\ref{tab:tab3}.

The second possible reason why the measured cross sections are larger than the
calculated ones from DF could be that the helium droplets are not spherical,
as assumed in the analysis, but have the shape of an oblate ellipsoid.  A
flattening of the droplets does not seem unfounded, due to the large angular
momentum of several $1000\:\hbar$ which some of the droplets may have as a
result of collisions with the residual gas before arriving in the scattering
region. A simple estimate indicates that in view of the large mass of the
droplet, the resulting rotation speed is slow compared with the colliding
secondary beam atoms. Thus the effect of a possible deformation on the
experimental results can be estimated from the calculated geometrical cross
section of a rotational ellipsoid averaged over all orientations.  In this
estimate vibrational distortions, which at the low temperature should only
have small amplitudes, are neglected.

The radius of the ellipsoid is assumed to have the length $a$ and the radius
of the axis of rotational symmetry is given by $b$. For $a\!>\!b$ one gets an
oblate and for $a\!<\!b$ a prolate ellipsoid.  The geometrical cross section
$\sigma_{\text{ell}}$ under an angle of view $\omega$ can be written as (see
Appendix~\ref{app:ellips}):
\begin{equation}
  \sigma_{\text{ell}} = \frac{4}{3}\: \pi\: a \: \sqrt{ \frac{b^4 + a^4 \tan^2
       \omega
      } {b^2 + a^2 \tan^2 \omega } } \: \sin \left[ \omega + \arctan \left(
  \frac{b^2 }{a^2 \tan \omega } \right)\right] \; .
\label{eq:sigmaell}
\end{equation}
The effective cross section is obtained by averaging numerically over all
orientations,
\begin{equation}
  \overline{\sigma_{\text{ell}}} = \frac{1}{2} \int_0^{\pi} d\omega \; 
      \sin(\omega) \; \sigma_{\text{ell}}(\omega) \; . 
\label{eq:meansigma}
\end{equation}
The resulting mean cross section is shown as a function of the ratio of the
radii $b/\!a$ in Fig.~\ref{fig:ellips}, scaled to the cross section of a
sphere with the same volume. For a real droplet with lower density at the
surface the effect will be even greater, since the surface area of the
ellipsoid is larger. To explain the difference of 5--10\% between the DF
results and the experiments, oblate shaped droplets with a value of $b/\!a
\approx 0.6$ or prolate droplets with $b/\!a \approx 2.5$ would be required.
Oblate shapes with such small average $b/\!a$ ratios seem unlikely in view
of the relatively small fraction ($\approx$15\%) of the incident beam droplets
which undergo collisions with the residual gas. Thus we conclude that even in
the very unlikely case that the droplets are somewhat distorted this has
little effect on the measured surface thickness.
 
%##########################################################################

\section{Conclusions}
\label{sec:conclusions}
By using a combination of scattering techniques (deflection and attenuation),
we have shown that it is possible to measure the average densities of large
Helium droplets. These results have been compared with density functional
calculations which have been carried out for the large droplets with $N=10^3$
to $10^4$ Helium atoms studied in the experiment. Overall agreement can be
considered to be satisfactory but on closer examination the predicted
effective integral cross sections are too small be about 10--20\% and the
densities too large by about 10--25\%, which is larger than the errors which
are estimated to be about 5\%. An attempt to fit the experimental data using a
symmetric model density profile yields a value for the 10--90\% thickness of
$t=6.4 \pm 1.3 $\AA\ which is larger than the density functional value of
5.7~\AA\ and is consistent with the differences in the densities.  If an
asymmetric profile is assumed 1--2\AA\ larger thicknesses are obtained which
would increase the discrepancy. The assumptions made in the analysis were
critically examined and an explanation for the small discrepancy cannot be
provided. As discussed in connection with Table~\ref{tab:tab3} the situation
is similar to the surface of the bulk where experiments also yield thicknesses
which are larger than predicted by most theories.

The experimental method has also been recently applied to
$^3$He droplets \cite{HT97} and there the average densities relative to the
bulk are even smaller than found here for $^4$He, a trend which is consistent
with recent Thomas-Fermi theoretical calculations \cite{B97}.

In the future it is conceivable that the present experimental method can be
further improved to provide a sensitive quantitative probe of the outer
surface region of large droplets.  Here it is interesting to speculate what
effect the large condensate fraction in this outer region \cite{GS95} would
have on the cross section. One can also explore the existence of a Landau
velocity below which the interaction should disappear, as it happens in the
bulk liquid at 57 m/s. Experiments under such conditions are now possible
\cite{HKT97} and are envisaged in the future.

\section*{Acknowledgements}

The authors are grateful to B. Schilling for his contribution to the early
phases of this work. They thank U. Henne and M. Lewerenz for many valuable
discussions and M. Faubel and B. Whaley for their critical reading of the
manuscript.

%##########################################################################
\appendix

\section{Integral Cross Section}
\label{app:cs}
The quantum mechanical differential cross section for elastic scattering from
the secondary beam gas atoms was calculated using a partial wave expansion
method \cite{EST88} in order to estimate the effect of the forward diffraction
contribution on our results.  The following spherical model potential obtained
by integrating the Lennard-Jones 12-6 potential between the scattering atom
and the He atoms of the droplet was assumed \cite{GV74}:
\begin{equation}
 V_N(r) = \frac{4 N \varepsilon \kappa^6  }{ \left(r^2 - R^2\right)^3 } \:
     \left[ \frac{\kappa^6 \left(r^6 +\frac{21}{5} r^4 R^2 + 3 r^2 R^4 +
                \frac{1}{3} R^6 \right) }    { \left(r^2 - R^2 \right)^6  }
        \: - 1 \right] \; ; \; r > R \; . 
\label{eq:pot_gs}
\end{equation}
The integration is simplified by assuming the He atoms to be homogeneously
distributed within a sphere of radius $R$. Here the effective radius $R$
corresponds to the radius $R_N=r_0N^{1/3}$ of a droplet with $N$ atoms reduced
by the effective radius $r_0$ of one atom in the same droplet, i.e., $R=R_N
-r_0=(N^{1\!/\!3}-1)r_0$. Estimates of the effective radius $r_0$ of one
helium atom in a droplet can be extracted from quantum many-body calculations.
For instance, Pandharipande et al. \cite{PZP83} reported the value $r_0(N) =
2.24 + 0.38 N^{-\frac{1}{3}} + 2.59 N^{-\frac{2}{3}}$.  The parameters 
$\varepsilon$ and $\kappa$ are such that, in the limiting case $N=1$,
the potential $V_N$ reduces to the two-body Lennard-Jones He-Kr or He-Ar
potential:
\begin{equation}
 V_1(r) = 4 \varepsilon \left[ (\kappa/r)^{12} - (\kappa/r)^6  \right]
   \: \equiv \: \varepsilon \left[ \left( \frac{R_m}{r}\right)^{12} -
    2\left(\frac{R_m}{r} \right)^6  \right] \; ,
\end{equation}
where $R_m=2^{1/6} \kappa$.  We use $\varepsilon=2.67$ meV and $R_m=3.70$ \AA\ 
for He-Kr, and $\varepsilon=2.59$ meV and $R_m=3.40$ \AA\ for He-Ar
\cite{DK86}.  Although the inner repulsive potential is not realistic for
central collisions with the strongly absorbing liquid core of the He droplet,
the potential Eq.~(\ref{eq:pot_gs}) should be a good approximation for
describing the glancing collisions, which are the relevant ones in determining
the elastic contribution to the integral cross section.

The integral cross section of real He droplets consists of the elastic cross
section $\sigma_e$ resulting from diffraction in large impact collisions and
the absorption cross section $\sigma_a$ in more central collisions. 
In the calculations with the hard core potential Eq.(\ref{eq:pot_gs}) the
angularly distributed isotropic part of the differential cross section takes
account  of the effect of the absorption cross section $\sigma_a$.  In our
case, where the de Broglie wavelength of relative motion $\lambda$ is much
smaller than the droplet radius, all particles with impact parameter smaller
than approximately $R_N$ are expected to be absorbed. The contribution of
particles with angular momentum $l$ to the cross section is given by
$\sigma_{l}=(2l+1) \pi\lambdabar^2$ \cite{BW79}. Then the
integral cross section is given approximately by:
\begin{equation}
 \sigma \; \simeq\;  
        \sum_{l=0}^{{R_N}/{\lambdabar}} (2l+1) \pi \lambdabar^2 
        \; + \;
        \sum_{l={R_N}/{\lambdabar}}^{\infty} \sigma_{_{el,l}} 
        \;\simeq\; \pi R_N^2 \;+\; \sigma_{e} . 
\end{equation}
The part $\sigma_e$ was calculated quantum mechanically exactly for the
potential Eq.~(\ref{eq:pot_gs}). Since the calculated elastic scattering is
found to be sharply peaked in the forward direction its contribution to the
integral cross section can be neglected for center of mass angles $\vartheta$
larger than about 5$^\circ$. The laboratory scattering angle $\Theta$ is in
the case of $m_{\text{drop}} \gg m_{\text{sec}}$ related to $\vartheta$ by 
\begin{equation}
  \Theta(\vartheta) = \frac{m_{\text{sec}}}{ v_{\text{drop}}(m_{\text{sec}}+ 
    m_{\text{drop}})} 
      \left\{ v_{\text{rel}}  \sin \left[ 
      \arcsin \left(\frac{v_{\text{sec}} \sin \beta}{v_{\text{rel}}}\right)+ 
         \vartheta \right] - v_{\text{sec}} \sin \beta 
         \right\} \; ,
\label{eq:cm2lab} 
\end{equation}
where $v_{\text{rel}}$ is the relative speed of the two beams and $\beta$ is
the angle between them. Since the mass $m_{\text{drop}}$ of the primary beam
droplets is much larger than the mass $m_{\text{sec}}$ of the secondary gas
atoms the angle $\Theta$ turns out to be very small. For example a center of
mass angle $\vartheta=5^\circ$ corresponds to laboratory angles $\Theta$ of
about $1$~mrad for droplets with 1000 atoms, which is comparable with the
angular resolution of our apparatus.

The theoretical differential cross sections were convoluted with the angular
distributions $S(\Theta)$ of the incident beam (beam profiles). These latter
angular distributions were measured with the same method as described in
Section \ref{sec:droplet_sizes} with the secondary beam blocked with a beam
flag located in the scattering chamber. In this arrangement the residual gas
pressure and its effect on the beam profile is the same as in the scattering
experiment. The predicted effective integral cross section is then given by:
\begin{equation}
  \sigma_{\text{app}}^{int} = 2 \pi \int^{\pi}_{\vartheta_{\text{app}}} 
  d \vartheta
  \sin(\vartheta) \left\{ \int_{- \pi}^{\pi} d \zeta \: 
  \left(\frac{d\sigma (\vartheta -\zeta)}{d\omega}\right)_{\text{th}}
  \: s(\zeta) \right\} , 
\end{equation}
where the $\vartheta_{\text{app}}$ represents the effective geometrical center
of mass resolution given by the slit in front of the scattering center and in
front of the detector and $s(\vartheta)$ is the measured angular profile of
the unscattered beam transformed into the center of mass system. The quantity
$(d\sigma/d\omega)_{\text{th}}$ is the calculated elastic differential cross
section.  Fortunately the resulting value of
$\sigma_{\text{app}}^{\text{int}}$ turns out to be very close to the classical
cross section $\sigma_{\text{class}}=\pi R_N^2 $; the two cross sections agree
to within 1\% for droplets with 1000 atoms and for larger droplets the
agreement is even better.  Thus the corrections discussed above justify the
assumption that the classical cross section is measured.

%##########################################################################

\section{Density functional}
\label{app:functional}

The explicit expression of the correlation energy $E_c$, entering
the density functional (\ref{eq:ec}), is given by \cite{DLP95,CDL96}:
\begin{eqnarray}
E_c [\rho]  &=&  \int \! d{\bf r} \Big\{
        {1\over 2} \int \!  d{\bf r}' \ \rho({\bf r}) V_l(|{\bf r}-
         {\bf r}'|) \rho({\bf r}')
        \ + \ {c_2 \over 2} \rho({\bf r})  (\overline{\rho_h} ({\bf r}))^2
        \ + \ {c_3  \over 3} \rho({\bf r}) (\overline{\rho_h} ({\bf r}))^3
\nonumber
\\     &-& {\hbar^2 \over 4m} \alpha_s \int \! d{\bf r}' \ F(| {\bf r}
                -{\bf r}' |)
               \left(1-{\rho({\bf r}) \over \rho_{0s}} \right)
               \nabla \rho({\bf r}) \cdot \nabla \rho({\bf r}')
               \left(1-{ \rho({\bf r}) \over \rho_{0s}} \right) \Big\}  \; .
\label{ec}
\end{eqnarray}
The two-body interaction $V_l$ is the Lennard-Jones interatomic potential,
with the standard parameters $\alpha =2.556$ \AA \ and $\varepsilon=10.22$ K,
screened at short distance ($V_l\equiv 0$ for $r<h$, with $h=2.1903$\AA). The
two terms with the parameters $c_2 = -2.411857 \times 10^4$ K \AA$^6$ and
$c_3=1.858496 \times 10^6$ K \AA$^9$ account phenomenologically for short
range correlations between atoms. The weighted density 
$\overline{\rho_h}({\bf r})$ is the average of $\rho$ over a sphere of 
radius $h$ centered in ${\bf r}$.  The last term, depending on the gradient
of the density in different points, is introduced to reproduce the static 
response function in the roton region. The function $F$ is a simple Gaussian, 
$F (r)= \pi^{-3/2} \ell^{-3} \exp(-r^2/\ell^2)$ with $\ell=1$~\AA, while 
$\alpha_s=54.31$~\AA$^3$ and $\rho_{0s} =0.04$ \AA$^{-3}$.

%##########################################################################

\section{Mean geometrical cross section of a rotational ellipsoid}
\label{app:ellips}

A rotational ellipsoid is assumed with the axis of rotational symmetry $b$
parallel to the z-axis and the other two radii $a$:
\begin{equation}
  \frac{x^2+y^2}{a^2}  +  \frac{z^2} {b^2} = 1 \; . 
\label{eq:para_ell}  
\end{equation}
This can be written in spherical coordinates:
\begin{equation}
  r(\Theta,\Phi)=\frac{b} {1 -\epsilon^2 \sin^2(\Theta)} \:; \hspace{1cm}
  \epsilon=\frac{\sqrt{a^2-b^2}}{a} < 1 \;.
\end{equation}
Since the visible cross section $\sigma_{\text{ell}}$ of the ellipsoid from a
viewing point with a polar-angle $\omega$ does not depend on the azimuth-angle
$\Phi$, the problem can be reduced to the x-z-plane. The visible cross section
$\sigma_{\text{ell}}$ is given by the projection of the cutting plane through
the ellipsoid which is defined by the tangents with angle $\omega$. Since
every cutting plane of an ellipsoid is an ellipse, the area of the cutting
plane $\sigma_s$ can written as:
\begin{equation}
  \sigma_s(\omega) = \pi a r_s(\omega) \;.
\label{eq:ss}
\end{equation}
The projection on a plane perpendicular to the viewing direction is given by:
\begin{eqnarray}
  \sigma_{\text{ell}}(\omega) & = & \sigma_s(\omega) \cos(\omega -
       \Theta_s) \;.
\label{eq:sell}
\end{eqnarray} 
To find the vector ${\bf r_s} = { r_s \choose {\pi - \Theta_s} } =
{x_s \choose z_s }$ lying in the cutting plane the equation of a
tangent at the ellipse has to be calculated
\begin{equation}
  1  =  \frac{x x_s}{a^2} + \frac{z z_s}{b^2} \\
\end{equation}
which yields   
\begin{equation}
  z   =  \frac{b^2}{z_s} - \frac{b^2 x_s}{a^2 z_s} x \;. 
\end{equation}
The tangent must be parallel to the viewing direction $\omega$
\begin{equation}
  \cot(\omega) = \frac{dz}{dx} = - \frac{b^2 x_s}{a^2 z_s} \; .
\end{equation}
With equation (\ref{eq:para_ell}) follows
\begin{equation}
  x_s^2 = \frac{a^4 \cot^2(\omega)} {b^2 + a^2 cot^2(\omega)} \; ,
\label{eq:xs}
\end{equation}
and for $z_s$ and $r_s$
\begin{equation}
  z_s^2 = \frac{b^4}{ b^2 + a^2 \cot^2(\omega)} \;,
\label{eq:zs}
\end{equation} 
\begin{equation}
  r_s^2 = x_s^2+z_s^2 =\frac{b^4+ a^4 
\cot^2(\omega)}{b^2 + a^2 \cot^2(\omega)} \; .
\end{equation}
From Eqs.~(\ref{eq:xs}) and (\ref{eq:zs}) $\Theta_s$  can be calculated:
\begin{equation}
  \tan(\Theta_s)   = \frac{z_s}{x_s} = \frac{b^2}{a^2 \cot(\omega) } \;.   
\end{equation} 
The visible cross section of an ellipsoid follows from Eqs.~(\ref{eq:ss}) and
(\ref{eq:sell})
\begin{equation}
  \sigma_{\text{ell}}(\omega) = \pi\: a \: \sqrt{ \frac{b^4 + a^4 
      cot^2(\omega)}
          {b^2 + a^2 \cot^2(\omega) } } \:\: \cos\left[ \omega - \arctan 
           \left(\frac{b^2 }{a^2 \cot(\omega) } \right)\right] \; .
\end{equation}
The mean cross section is derived be integration over $\omega$ 
\begin{equation}
  \overline{\sigma_{\text{ell}}(\omega)} = \frac{1}{2}\int_0^{\pi} d\omega \;
         \sin(\omega) \; \sigma_{\text{ell}}(\omega) \; .
\end{equation} 

%############################################################################

%#########################################################################
\widetext
\begin{table*}[ht]
\caption{
  Experimental results as function of source temperature ($T_0$), source
  pressure ($P_0$) and secondary gas. The mean number of atoms $\overline N$
  and the half-width $\Delta N_{\frac{1}{2}}$ are the results of fitting the
  measured mass distributions from the deflection experiment with a
  log-normal distribution (parameters $\delta$ and $\mu$, see 
  Eq.~(\protect\ref{eq:lognorm})). 
  The mean classical integral cross section $\overline{\sigma}$ is obtained
  by attenuation of the droplet beam with the secondary beam. From this data
  the mean density of the droplets $\overline{\rho}$ as a fraction of the
  known bulk density ($\rho_{\text{bulk}}=0.0218~$\AA$^{-3}$) is obtained 
  directly. 
}

\begin{tabular}{  r  r  r  r  r  r  r  r  r }
$T_0$ [K]   &   $P_0$ [bar]   &   sec. gas   &   $\overline{N}$ & 
        $\Delta N_{\frac{1}{2}} $   &    $\delta$    &   $\mu$   &
        $\overline{\sigma}$\ [\AA$^2$]   &   
        $\overline{\rho} / \rho_{\text{bulk}}$ \\
\hline
24.0  & 40 & Kr &703     & 667  & 0.426& 6.46  &2266   & 0.40 \\
22.0  & 40 & Kr &1700    & 1632 & 0.407& 7.36  &3138   & 0.59 \\
20.0  & 40 & Kr &2617    & 2373 & 0.528& 7.73  &4519   & 0.53 \\  
18.0  & 40 & Kr &4700    & 4158 & 0.573& 8.29  &6259   & 0.58 \\       
17.0  & 40 & Kr &6130    & 5331 & 0.603& 8.54  &7108   & 0.62 \\      
16.0  & 40 & Kr &7741    & 6484 & 0.662& 8.74  &7661   & 0.70 \\          
15.0  & 40 & Kr &8900    & 7719 & 0.607& 8.91  &8540   & 0.69 \\         
13.5  & 40 & Kr &13000   & 11240& 0.612& 9.29  &9538   & 0.85 \\          
26.0 \tablenotemark[1]
      & 80 & Kr &1460    & 1298 & 0.565& 7.13  &3106   & 0.51 \\ 
24.0 \tablenotemark[1]
      & 80 & Kr &2700    & 2524 & 0.468& 7.79  &4111   & 0.62 \\
18.0 \tablenotemark[1]
      & 80 & Kr &5260    & 4374 & 0.673& 8.34  &6270   & 0.65 \\         
26.0  & 80 & Ar &2114    & 1967 & 0.478& 7.54  &3431   & 0.64 \\          
24.0  & 80 & Ar &3103    & 2835 & 0.514& 7.91  &4443   & 0.64 \\
20.0  & 80 & Ar &6458    & 5916 & 0.509& 8.64  &6615   & 0.70 \\
18.0  & 80 & Ar &9487    & 8530 & 0.544& 9.01  &9025   & 0.67 \\
\end{tabular}
\tablenotetext[1]{ The smaller cluster number sizes found with Kr instead of
 Ar as scattering gas at $P_0=80~bar$ is a consequence of a smaller nozzle
 with an estimated diameter of about $4\mu m$.}
\label{tab:exp_tab}
\end{table*}

%##########################################################################
\narrowtext
\begin{table}[ht]
\caption{
  The effective radii
  $R_{\text{eff}}=\protect \sqrt{\sigma_{\text{eff}} / \protect\pi}$ and 
  cross sections 
  $\sigma_{\text{eff}}$ obtained using  Eq.~(\protect \ref{eq:trans}) and
  DF calculations of density profiles. The effective density 
  $\protect \overline{\rho}$ as a fraction of the known bulk density 
  ($\rho_{\text{bulk}}=0.0218~$\AA$^{-3}$) and the 10--90\% surface 
  thickness $t$ are listed. 
} 

\begin{tabular}{ r  r  r  r  r }
$N$ &$R_{\text{eff}}$ [\AA] &$\sigma_{\text{eff}}$[\AA$^2$]& 
                $\overline{\rho} / \rho_{\text{bulk}}$ & $t$ [\AA] \\
\hline
1000    &25.75  &2083   & 0.64  & 5.6 \\
2000    &31.59  &3135   & 0.69  & 5.6 \\
3000    &35.69  &4000   & 0.72  & 5.7 \\
4000    &38.94  &4763   & 0.74  & 5.7 \\
5000    &41.69  &5459   & 0.76  & 5.7 \\
6000    &44.09  &6106   & 0.77  & 5.7 \\
7000    &46.22  &6713   & 0.78  & 5.7 \\
8000    &48.18  &7292   & 0.78  & 5.7 \\
9000    &49.97  &7843   & 0.79  & 5.7 \\
10000   &51.63  &8373   & 0.80  & 5.7 \\
\end{tabular}
\label{tab:tab_dfc}
\end{table}

%##########################################################################
\mediumtext
\begin{table*}[ht]
\caption{
  The 10--90\% surface thickness of $^4$He$_{\rm N}$ droplets and 
  bulk liquid helium from different published experimental and theoretical 
  works, where $N$ is the number of atoms in the droplets and $T$ 
  the temperature (the values for $N=\infty$ refer to the planar free 
  surface).
}

\begin{tabular}{ l  r  l  r  r }
& $N$   & method   &   $T$[K] & $t$[\AA]  \\
\hline
Osborne (1989) \cite{O89}        & $\infty$ & ellipsometry & 1.8 & 9.4 \\
\hline
Lurio et al. (1992) \cite{LRP92} &$\infty$ & X-Ray         & 1.13& 9.2 \\
                                 &         & extrapol.     &   0 & 7.6 \\
\hline
Pandharipande et al. (1983) \cite{PZP83}& 
                                  $< 728$  & GFMC,VMC      & 0  & 5.5--7.2 \\
\hline
Stringari, Treiner (1987) \cite{ST87}&
                                  $< 728$  & DF            & 0  & 8.8--9.2 \\
                                 &$\infty$ & DF            & 0  & 7 \\       
\hline
Sindzingre, Klein, Ceperley (1989) \cite{SKC89} &
                                   64-128  & PIMC    &0.5--2&$\simeq 6$\\
\hline
Guirao, Centelles, Barranco et al. (1992) \cite{GCB92} & 
                                 $\infty$  & DF      &0--4 & 6.5 (at 0.4~K) \\
\hline
Chin, Krotscheck (1995) \cite{CK95}& 20--112 & DMC        & 0  & $\simeq 6$ \\
                                   & 20--1000& HNC        & 0  & $\simeq 6$ \\
\hline
Barnett, Whaley (1995) \cite{BW95}& $<112$ & DMC          & 0  & $\simeq 6$ \\
\end{tabular}
\label{tab:tab3}
\end{table*}

%##########################################################################

\begin{figure}[ht]
%\begin{center}  
%\leavevmode
%\epsfysize=150mm \epsfbox{fig1.eps}
%\end{center}
\caption{Three typical measured angular distributions for a source pressure 
  of $P_0=80~bar$ and source temperatures of $T_0=25~K$ (a), $20~K$ (b) 
  and $17~K$ (c). Krypton was used as secondary beam gas. The signals with
  (filled circles) and without (open circles) a flag in front of the 
  secondary beam are show on a logarithmic scale. The weighted differences 
  of the two signals (diamonds) with the standard deviations are shown on
  a linear scale. The mean number of atoms in the droplets are: 
  a)~$\overline{N}=2602$, b)~$\overline{N}=6174$, c)~$\overline{N}=9834$. 
  The integral cross section is determined from the attenuation of the forward
  peak, ie. $0~mrad$. }
\label{fig:distr}
\end{figure}

%##########################################################################

\begin{figure}[ht]
%\begin{center}  
%\leavevmode
%\epsfxsize=\hsize \epsfbox{fig2.eps}
%\end{center}
\caption{The classical integral cross sections averaged over the measured
  number size
  distributions as functions of mean number of atoms $\overline{N}$. The empty
  symbols show the experimental results.  The solid line with filled diamonds
  is the result of the DF calculations. For comparison, the mean classical
  cross sections of spherical droplets with constant density are indicated as
  dashed lines for different values of the density
  ($\rho_{\text{bulk}}=0.0218~$\AA$^{-3}$). The different empty 
  symbols indicate
  the different experimental parameters: triangle: $P_0=40~bar$, Sec.-gas=Kr;
  square: $P_0=80~bar$, Sec.-gas=Kr; circle: $P_0=80~bar$, Sec.-gas = Ar.}
\label{fig:cs}
\end{figure}

%##########################################################################

\begin{figure}[ht]
%\begin{center}
%\leavevmode
%\epsfysize=120mm \epsfbox{fig3.eps}
%\end{center}
\caption{The density distributions a) calculated with a density 
  functional method for droplets with between $N=10^3$ and $N=10^4$ atoms with
  steps of $10^3$.  The corresponding transmission for a beam of krypton or
  argon atoms is shown in b). The effective radius $R_{\text{eff}}$ for a 
  droplet with $10^4$ atoms is also indicated. }
\label{fig:profiles}
\end{figure}

%##########################################################################

\begin{figure}[ht]
\caption{The experimental 10--90\% surface thickness of He droplets as
  function of the mean number of atoms $\overline{N}$ assuming the symmetric
  density profile (Eq.~(\protect\ref{eq:rho_r})) shown in the inset.  The mean
  value of $t$ is $6.4 \pm 1.3$\AA. The DF values are shown as a dotted line.
  The different empty symbols
  indicate the different experimental parameters: triangle: $P_0=40~bar$,
  Sec.-gas=Kr; square: $P_0=80~bar$, Sec.-gas=Kr; circle: $P_0=80~bar$,
  Sec.-gas = Ar. }
\label{fig:exp_prof}
\end{figure}

%##########################################################################

\begin{figure}[ht]
\caption{The effective integral  cross section of an ellipsoid averaged over 
  all possible orientations in units of the cross section of a sphere with the
  same volume. The radius $b$ is the axis of symmetry and $a$ indicates the
  other radius.}  
\label{fig:ellips} 
\end{figure}
%##########################################################################

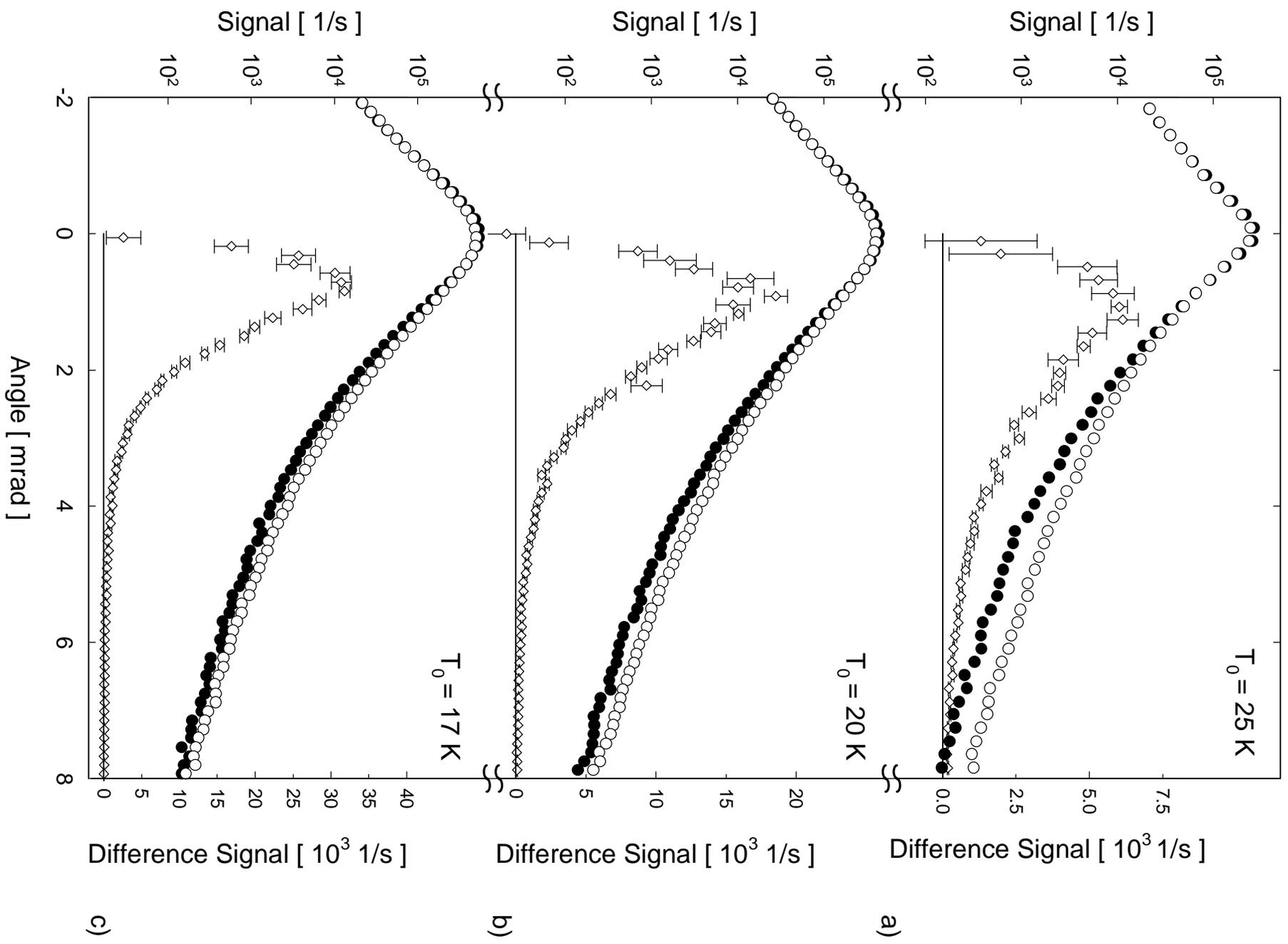
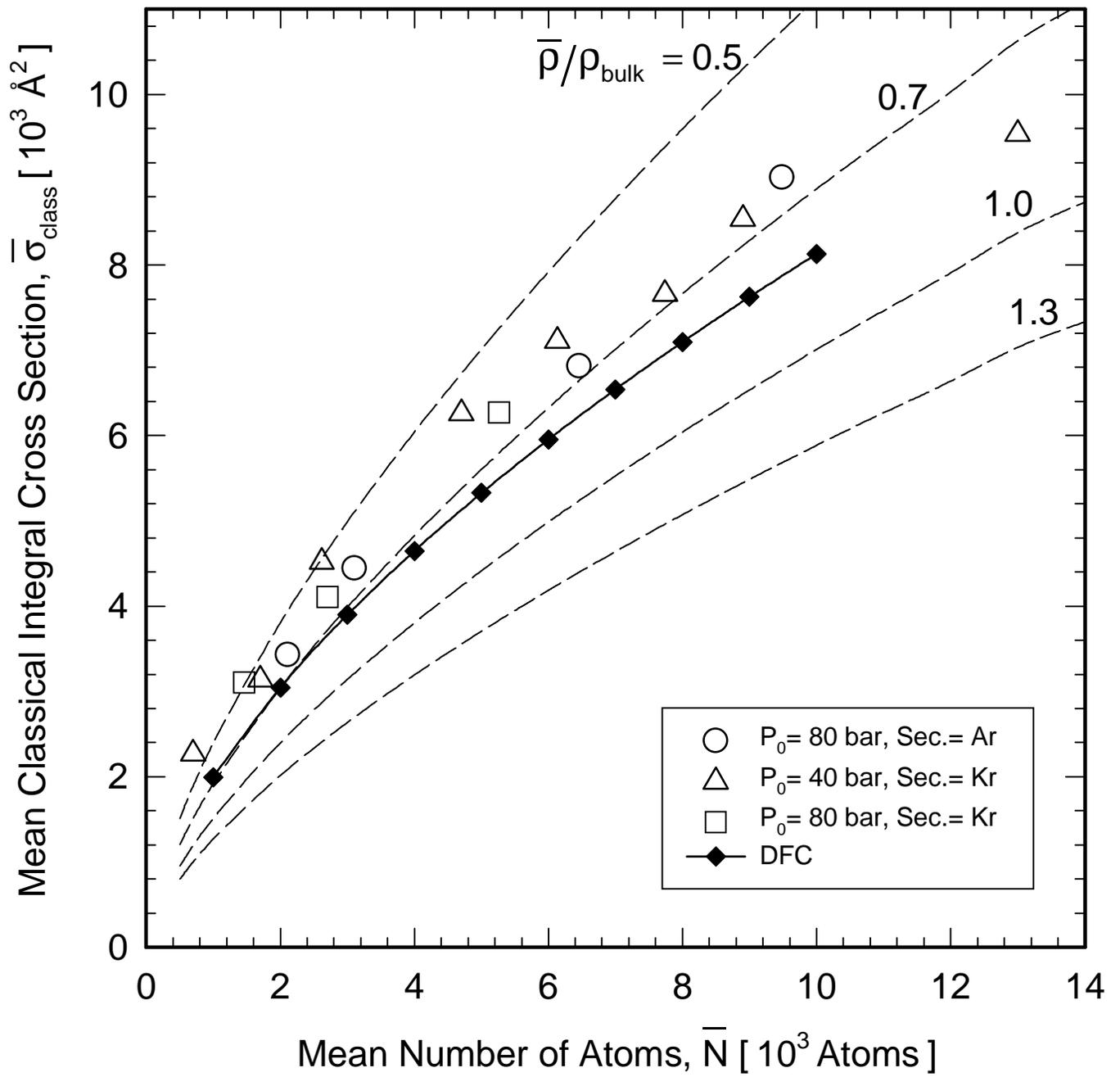
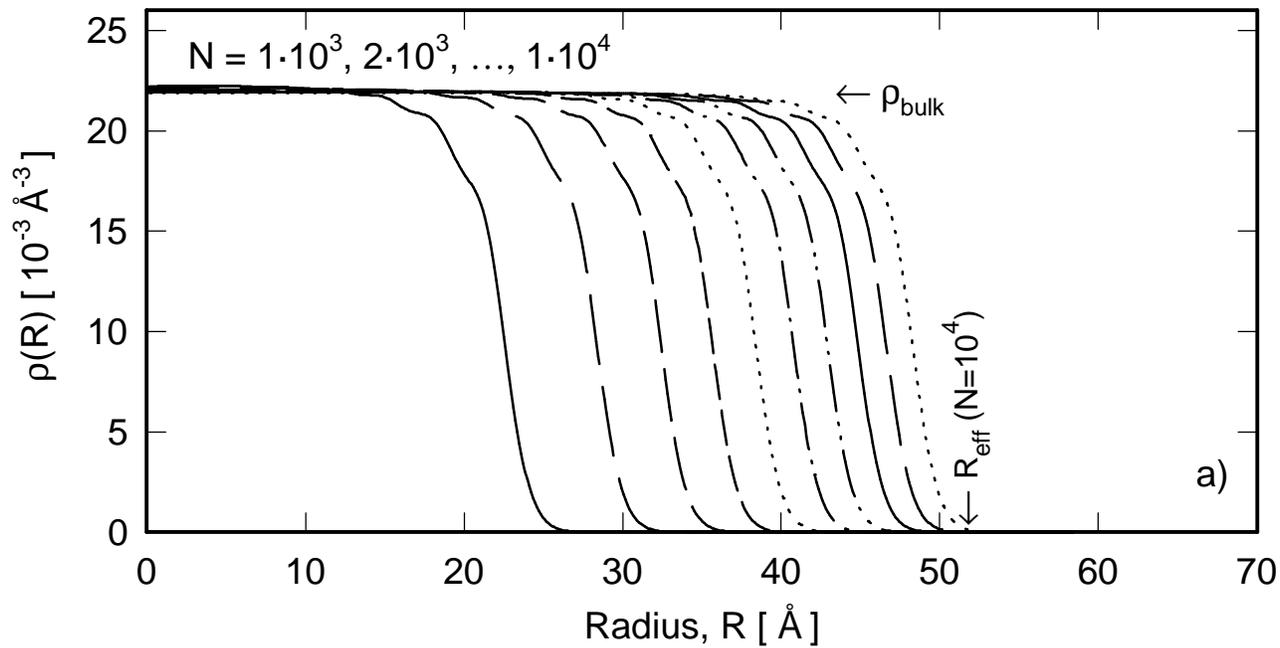
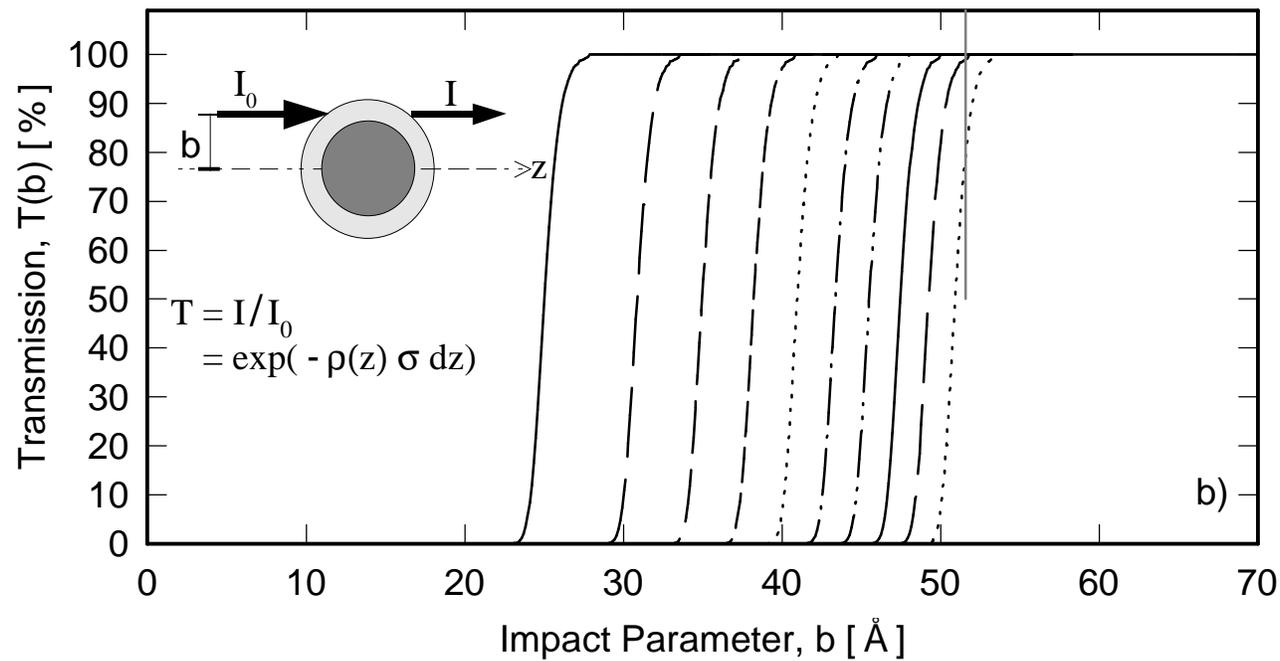
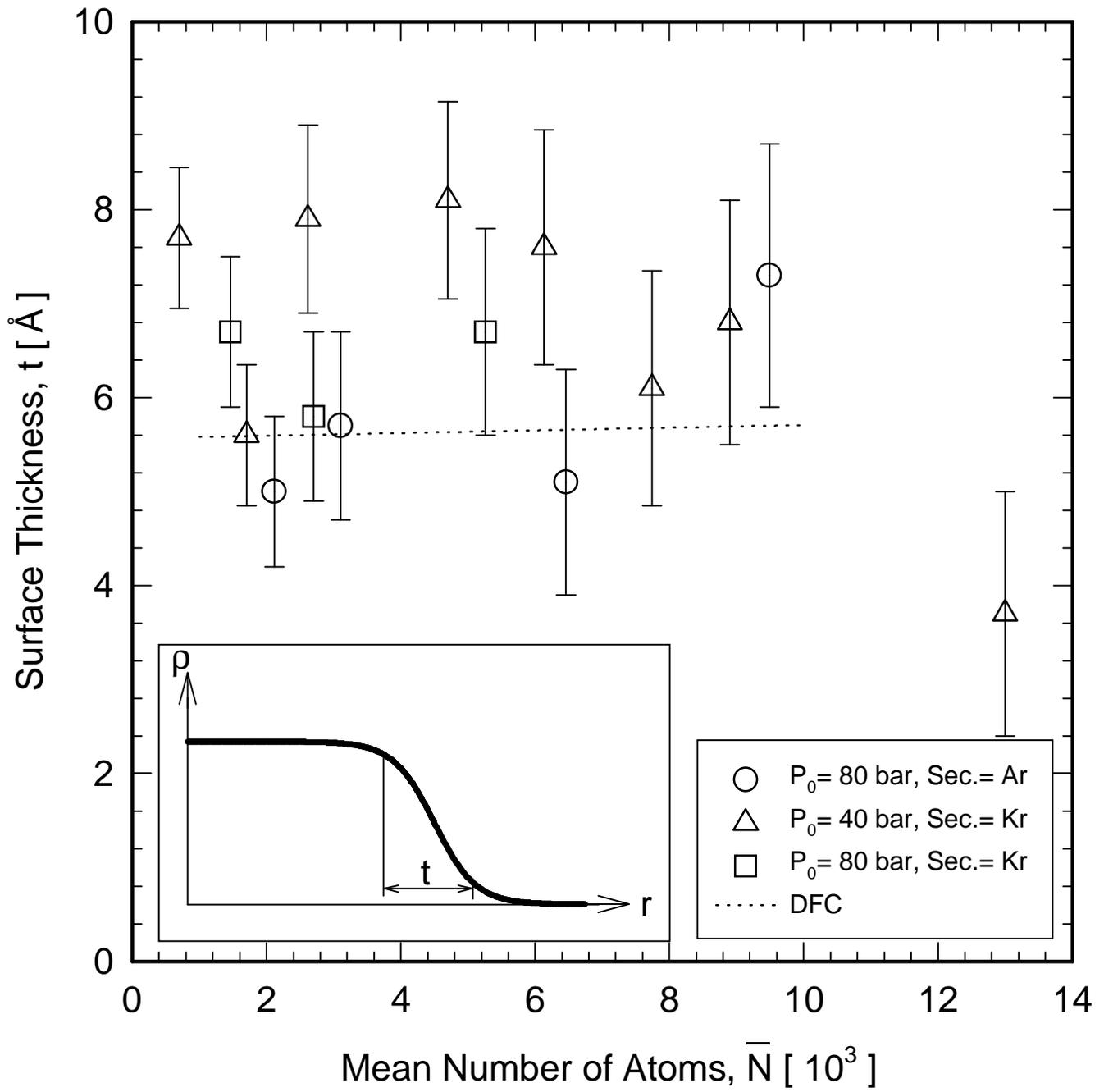
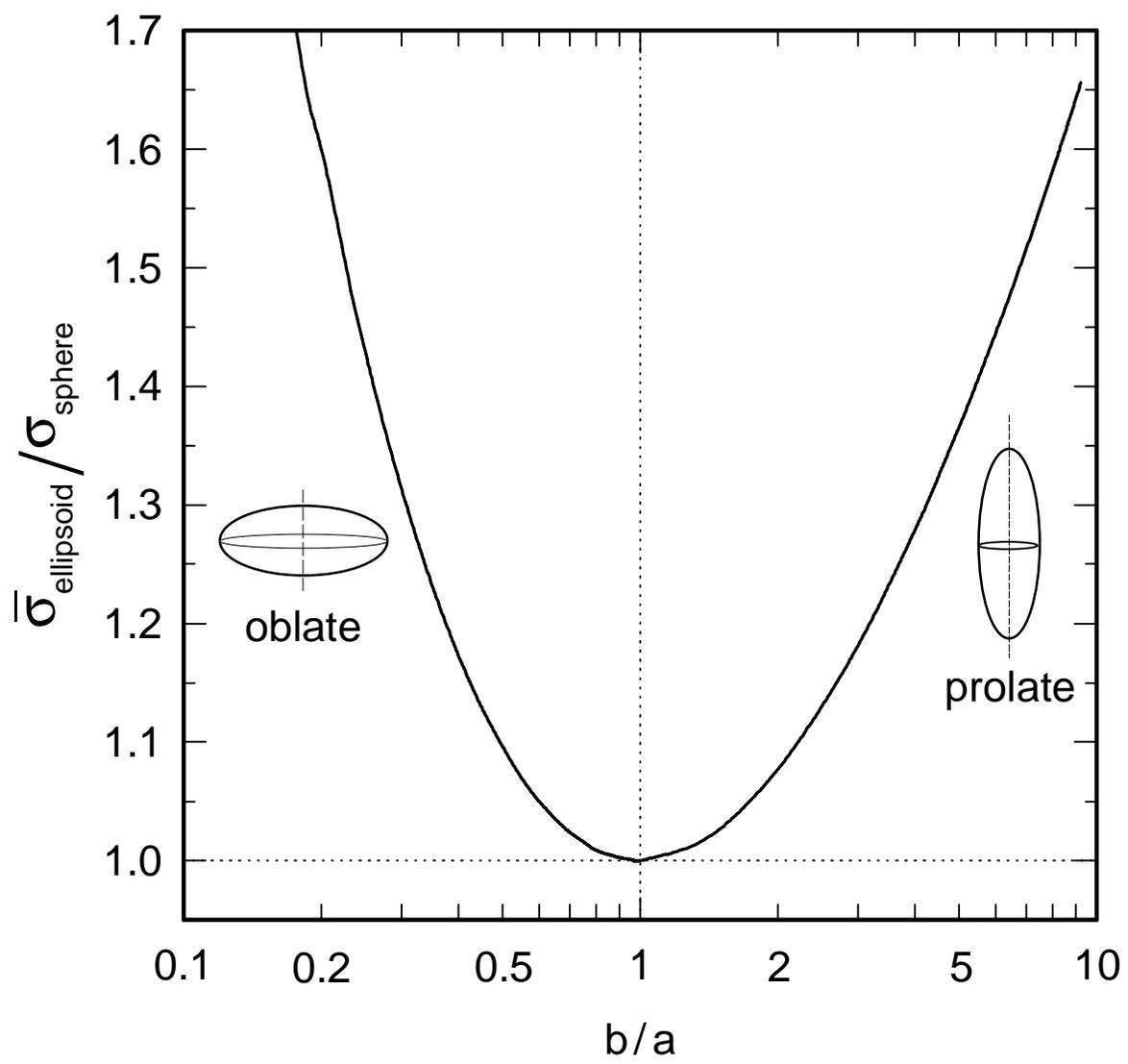
\end{document}